\documentclass[aps,pre,10pt,twocolumn]{revtex4-1}
\usepackage{graphicx}
\usepackage{epstopdf}
\usepackage{amsmath}
\usepackage{amssymb}
\usepackage{default}
\usepackage{natbib}
\newcommand{\be}{\begin{equation}}
\newcommand{\bea}{\begin{eqnarray}}
\newcommand{\bc}{\begin{center}}            
\newcommand{\ee}{\end{equation}}
\newcommand{\eea}{\end{eqnarray}}
\newcommand{\ec}{\end{center}}

\newcommand{\baa}{\begin{eqnarray*}}
\newcommand{\eaa}{\end{eqnarray*}}
\begin{document}
\title{Three-level laser heat engine at optimal performance with ecological function}
\author{Varinder Singh}
\email{varindersingh@iisermohali.ac.in}
\author{Ramandeep S. Johal}
\email{rsjohal@iisermohali.ac.in}
\affiliation{ Department of Physical Sciences, \\ 
Indian Institute of Science Education and Research Mohali,
Sector 81, S.A.S. Nagar, Manauli PO 140306, Punjab, India}
\begin{abstract}
Although classical and quantum heat engines work on entirely different fundamental principles, 
there is an underlying similarity. For instance, the form of efficiency at optimal performance may be similar for both types of engines. 
In this work, we study a three-level laser quantum heat engine operating at 
maximum ecological function (EF) which 
represents a compromise between the power output and the loss of power due to  entropy production. 
We derive analytic expressions for efficiency under the assumptions of  
strong matter-field coupling and high bath temperatures.
Upper and lower 
bounds on the efficiency exist in case of extreme asymmetric dissipation when the ratio of 
system-bath coupling constants at the hot and the cold contacts 
respectively approaches, zero or 
infinity. These bounds have been established previously 
for various classical models of Carnot-like engines.
We conclude that while the engine  
produces at least 75\% of the power output as compared with the maximum power conditions,
the fractional loss of power is appreciably low in case of the engine operating at maximum EF, thus making this objective function  relevant from an  environmental point of view. 
\end{abstract} 
%\keywords{quantum brownian motion}
\maketitle
\section{Introduction}
With the explosion of interest in quantum thermodynamics \cite{SV2016,Xuereb,KosloffEntropy}, we may be entering an era whereby energy conversion devices are able to harness non-classical properties like
coherence between internal states, entanglement, quantum degeneracy and so on. Thus, it is of great importance to ascertain the extent up to which these devices may surpass the performance of  
macroscopic, classical heat devices. 

On the other hand, rising concerns about the effects of human activity on the environment make it prudent that the new technologies be better from an ecological point of view. 
Most comparisons that are usually studied between quantum \cite{Scully2001, Kieu2004, Quan2007,Sir2008, GeorgeSir2011, Abe2011,
AgarwalChaturvedi2013,Campo2014, OzgurCorrelation, Salamon2016,Suman1,VenuSir2017, 
Bijay2017,George2018} and classical models of heat engines
\cite{Curzon1975,Esposito2010,IzumidaOkuda}, focus on the  optimization of power 
output \cite{Esposito2010,Esposito,Lutz2012,WangHe2013, Campo2016,  Campisi2016, Campo2017, Erdman2017, VascoCavina, Dorfman2018}. 
However, to be ecologically aware, we must 
care about the extent of entropy production which ultimately impacts the environment. As has been noted \cite{Chen2001,devosbook}, real thermal plants and practical heat engines may not operate at
maximum power point, but rather in a regime with a slightly smaller power output and appreciably larger efficiency. 
In recent years, a few such alternate measures of performance 
have been studied. Thus, the ecological function \citep{ABrown1991,VarinderJohal} or Omega function \cite{Hernandez2001}
and efficient-power function \cite{Yilmaz,VJ2018} fall under such a category, as they pay equal 
attention to both power and efficiency. 

In this work, we study the optimization of ecological function in the performance of a
three-level steady state laser heat engine \cite{Scovil1959}.
 The ecological function (EF)  is defined as \citep{ABrown1991}
 \be
E=P-T_c \dot{S}_{\rm tot},
\label{eco1}
\ee
where $P$ is the power output, $T_c$ is the temperature of the cold reservoir and 
$\dot{S}_{\rm tot}$ is the total rate of entropy production. Optimization of EF represents a compromise
between the power output and the loss of power due to entropy production. In the context of classical models, this function suggests optimal working conditions which lead to a drop of about 20\% in power output (compared to maximum power output), but on the other hand, reduce the entropy production by about 70\% \citep{ABrown1991}.  

Our second motivation for this analysis is to study the 
correspondence between classical and quantum heat engines (QHEs).
In most of the studies so far, QHEs show exotic behavior 
owing to additional resources such as quantum coherence
\cite{OzgurSuperradiant,OzgurQuantumFuel,Scully2011,Dorfman2018,Harbola2013, VenuSir2017, Petruccione2018, UzdinX}, 
quantum entanglement \cite{Zhang2007,Zhang2008,WangLiuHe2009,Hovhannisyan, Chiara2018}, squeezed baths \cite{Huang2012,LutzPRL,Alicki2015}, among others. 
Otherwise, QHEs may show a remarkable similarity to macroscopic heat engines. In such cases, 
Carnot efficiency provides an upper bound on the efficiencies of QHEs 
operating between two heat reservoirs. The irreversible 
operation of quantum engines with finite power output  has many similarities to macroscopic endoreversible 
engines and the low-dissipation model \cite{Esposito2010, Esposito}. 
Also in the high temperatures limit, QHEs are expected to behave like classical heat engines \cite{Geva1992,GevaKosloff1992}. We confirm these expectations in the analysis of the 
three-level laser engine using the ecological function.

The paper is organised as follows. In Sec.II, we discuss the model of three-level laser quantum heat engine. In Sec. III, 
we obtain the general expression for the efficiency of the engine operating at maximum EF and find lower and upper bounds 
on the efficiency for two different optimization schemes. In Secs. IV and V, we compare the performance of heat engine
operating at maximum EF to the engine operating at maximum power. We conclude in Sec. VI by highlighting the key results.

\begin{figure}   
 \begin{center}
\includegraphics[width=8.6cm]{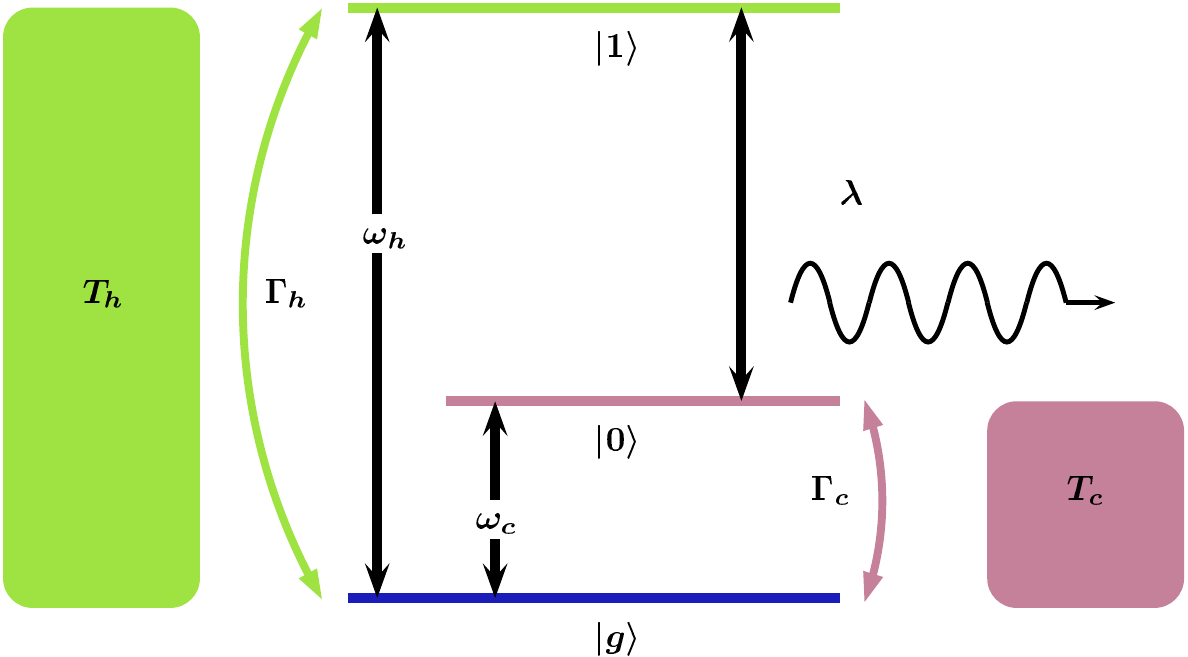}
 \end{center}
\caption{(Color online) Model of three-level laser heat engine continuously coupled to two reservoirs of temperatures
$T_h$ and $T_c$ having coupling constants $\Gamma_h$ and $\Gamma_c$, respectively. The system is interacting with a
classical single mode field. $\lambda$ represents the strength of matter-field coupling.}
\end{figure}

\section{Model of Three-Level Laser Quantum Heat Engine}
One of the simplest QHEs is three-level laser heat engine \cite{Scovil1959} introduced by Scovil and 
Schulz-Dubois (SSD). It converts the incoherent thermal energy of 
heat reservoirs into a coherent laser output. The model has been studied extensively in the literature, and three-level systems are also employed to study quantum absorption refrigerators \cite{Alonso2014,Alonso2014B,Segal2018,Scarani}. The model 
proposed by Scovil and DuBois was further analyzed by Geva and Kosloff \cite{Geva1994,Geva1996} in the spirit of finite time 
thermodynamics. In the presence of strong time dependent external fields, they optimized the power output 
of the amplifier w.r.t  different control parameters. In their model, the second law of thermodynamics 
is generally satisfied if one incorporates the effect of external field on the dissipative superoperator.
In a series of papers \cite{BoukobzaTannor2006A,BoukobzaTannor2006B,BoukobzaTannor2007}, 
Boukobza and Tannor formulated a new way to partition energy into heat and work \cite{Alicki1979}. They 
applied their analysis to  a three level amplifier continuously coupled to two reservoirs and  to a classical
single mode field \cite{BoukobzaTannor2007}. Their formulation is quite general and one does not need to incorporate the effect of external 
field on the dissipative term of the Liouvillian, and yet the second law of thermodynamics is always satisfied
at the steady state. In this paper, we use the formalism of Ref. \cite{BoukobzaTannor2007} to study the optimal performance of a 
three-level QHE operating in high temperature regime.

More precisely, the model consists of a three-level system continuously coupled to two thermal reservoirs and to a single 
mode classical field (see Fig. 1). A hot reservoir at temperature $T_h$ drives the transition between the ground level
$\vert g\rangle$ and top level $\vert 1\rangle$, whereas the transition between the intermediate level 
$\vert 0\rangle$ and ground level $\vert g\rangle$ is  constantly de-excited by a cold reservoir at 
temperature $T_c$. The power output mechanism is modeled by coupling the levels $\vert 0\rangle$ and
$\vert 1\rangle$ to a classical single mode field. The Hamiltonian of the system is given by: 
$H_0=\hbar \sum \omega_k \vert k\rangle\langle k\vert$ where the summation runs over all three states and 
$\omega_k$ represents the relevant atomic frequency. The interaction with the single mode lasing field of frequency 
$\omega$, under the rotating wave approximation, is described by the semiclassical hamiltonian: 
$V(t)=\hbar \lambda (e^{i\omega t} \vert 1\rangle\langle 0\vert + e^{-i\omega t} \vert 0\rangle\langle 1\vert)$; 
$\lambda$ is the matter-field coupling constant. The time  evolution of the system is described by the following 
master equation:
\begin{equation}
\dot{\rho} = -\frac{i}{\hbar} [H_0+V(t),\rho] + \mathcal{L}_{h}[\rho] + \mathcal{L}_{c}[\rho],
\end{equation}
where $\mathcal{L}_{h(c)}$ represents the dissipative Lindblad superoperator describing the system-bath 
interaction with the hot (cold) reservoir:
\begin{eqnarray}
\mathcal{L}_h[\rho] &=& \Gamma_h(n_h+1)(2\vert g\rangle\langle g\vert \rho_{11} - \vert 1\rangle\langle 1\vert\rho
- 
\rho\vert 1\rangle\langle 1\vert) \nonumber
\\
&& +\Gamma_h n_h (2\vert 1\rangle\langle 1\vert\rho_{gg}-\vert g\rangle\langle g\vert\rho
-
\rho\vert g\rangle\langle g\vert),\label{dissipator1}
\end{eqnarray} 
\begin{eqnarray}
\mathcal{L}_c[\rho] &=& \Gamma_c(n_c+1)(2\vert g\rangle\langle g\vert \rho_{00} - \vert 0\rangle\langle 0\vert\rho
- 
\rho\vert 0\rangle\langle 0\vert) \nonumber
\\
&& +\Gamma_c n_c (2\vert 0\rangle\langle 0\vert\rho_{gg}-\vert g\rangle\langle g\vert\rho
-
\rho\vert g\rangle\langle g\vert).\label{dissipator2}
\end{eqnarray}
Here $\Gamma_h$ and $\Gamma_c$ are the Weisskopf-Wigner decay constants, and 
$n_{h(c)}= 1/(\exp[\hbar\omega_{h(c)}/k_B T_{h(c)}]-1)$ is average occupation number of photons 
in hot (cold) reservoir satisfying the relations $\omega_c=\omega_0-\omega_g$, $\omega_h=\omega_1-\omega_g$.

In this model, it is possible to find a rotating frame in which the steady-state density matrix $\rho_R$
is time independent \cite{BoukobzaTannor2007}. Defining $\bar{H}=\hbar (\omega_g \vert g\rangle\langle g\vert 
+ \frac{\omega}{2} \vert 1\rangle\langle 1\vert - \frac{\omega}{2} \vert 0\rangle\langle 0\vert ) $, 
an arbitrary operator $A$ in the rotating frame is given by $A_R=e^{i\bar{H}t/\hbar}Ae^{-i\bar{H}t/\hbar}$. 
It can be seen that $\mathcal{L}_h[\rho]$ and $\mathcal{L}_c[\rho]$ remain unchanged under this transformation.
Time evolution of the system density matrix in the rotating frame can be written as
\begin{equation}
\dot{\rho_R} = -\frac{i}{\hbar}[H_0-\bar{H}+V_R,\rho_R] + \mathcal{L}_h[\rho_R] + \mathcal{L}_c[\rho_R]\label{dm1}
\end{equation}
where $V_R=\hbar\lambda(\vert 1\rangle\langle 0\vert + \vert 0\rangle\langle 1\vert)$.

For a weak system-bath coupling, the output power, the heat flux and the  efficiency of the engine can be defined \cite{BoukobzaTannor2007}, as follows:
\begin{eqnarray}
P &=& \frac{i}{\hbar} {\rm Tr}([H_0,V_R]\rho_R), \label{power1} \\
\dot{Q_h} &=&  {\rm Tr}(\mathcal{L}_h[\rho_R]H_0), \label{heat1} \\
\eta &=& \frac{P}{\dot{Q_h}}. \label{efficiency0}
\end{eqnarray}
Plugging the expressions for $H_0$, $V_R$ and $\mathcal{L}_h[\rho_R]$, and calculating the traces (see Appendix A) appearing 
on the right hand side of Eqs. (\ref{power1}) and (\ref{heat1}), the power and heat flux can be written as:
\begin{eqnarray}
P &=&i\hbar\lambda(\omega_h-\omega_c)(\rho_{01}-\rho_{10}), \label{power2} \\
\dot{Q_h} &=&  i\hbar\lambda\omega_h (\rho_{01}-\rho_{10}),\label{heat2}
\end{eqnarray}
where $\rho_{01} = \langle 0\vert\rho_R\vert 1\rangle$ and $\rho_{10} = \langle 1\vert \rho_R\vert 0\rangle$.
Then, the efficiency is given by
\begin{equation}
\eta = 1 - \frac{\omega_c}{\omega_h}. \label{efficiency}
\end{equation}
From Eq.(\ref{power6}), the positive power production condition implies that $\omega_c/\omega_h\geq T_c/T_h$.
Hence $\eta \leq \eta_{\rm C}^{}$.
\section{Optimization of Ecological Function }
The optimal performance of SSD 
engine at maximum power has already been studied recently \cite{Dorfman2018}. 
In this work, we optimize the EF which represents a trade-off between 
power output and loss of power in the system. 
We identify the total rate of entropy production in the heat reservoirs
due to operation of our engine as
\begin{equation}
\dot{S}_{\rm tot} = \frac{\dot{Q_c}}{ T_c}-\frac{\dot{Q_h}}{ T_h}.
\label{entropyproduction}
\end{equation}
In the steady state, the entropy of the system remains constant.
Substituting Eq. (\ref{entropyproduction}) in (\ref{eco1}), the EF can be written as
\begin{equation}
E = 2P-(1-\tau) \dot{Q_h}, \label{eco2}
\end{equation}
where $\tau =T_c/T_h$.
Using Eqs. (\ref{power2}) and (\ref{heat2}), we recast Eq. (\ref{eco2}) as 
\begin{equation}
E = i\hbar\lambda(\rho_{01}-\rho_{10})[2(\omega_h-\omega_c) - (1-\tau)\omega_h]. \label{eco4}
\end{equation}
\begin{figure}   
 \begin{center}
\includegraphics[width=8.6cm]{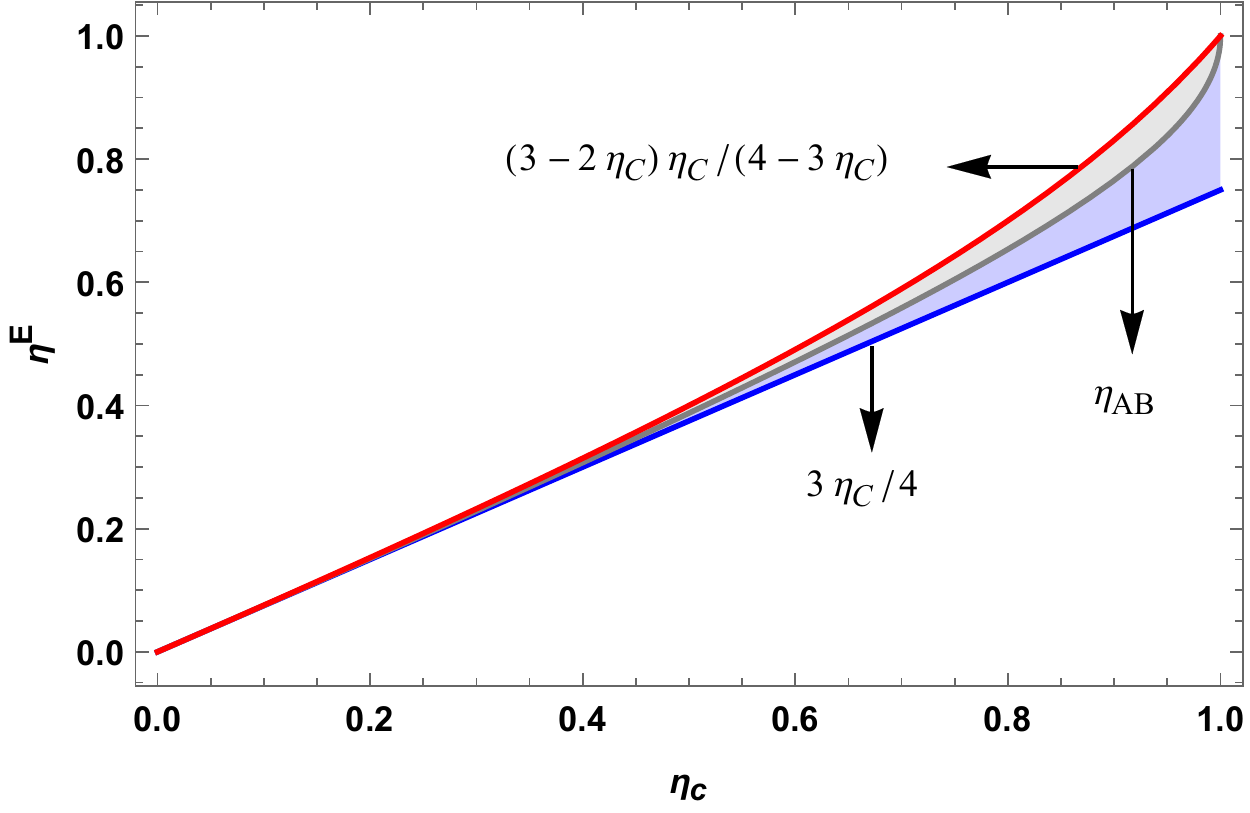}
 \end{center}
\caption{(Color online)  Efficiency $\eta^E$ versus the Carnot efficiency $\eta_{\rm C}^{}$ for the SSD engine. 
$\eta_{\rm AB}$ serves as the upper bound for the case with fixed $\omega_h$ and the lower bound for  a fixed $\omega_c$.}
\end{figure}
Now we optimize $E$   w.r.t. the transition frequencies $\omega_h$ and $\omega_c$, and
then calculate the corresponding efficiency at maximum ecological function (EMEF). In order to obtain 
analytic expressions in a closed form for the EMEF, we will work in the high temperatures regime and
assume the matter-field coupling to be very strong as compared to the system-bath coupling ($\lambda \gg   \Gamma_{h,c}$).
In the high temperatures limit, we set $n_h\simeq k_B T_h/\hbar\omega_h$ and $n_c\simeq k_B T_c/\hbar\omega_c$. 
The function $E$ can then be written in the following form (Appendix B)
\begin{equation}
E \simeq \frac{2\hbar \Gamma_h(\omega_c - \tau\omega_h)[2(\omega_h-\omega_c) - (1-\tau)\omega_h]}
 {3(\omega_c \gamma + \tau\omega_h )}, \label{ef}
\end{equation}
where $\gamma=\Gamma_h/\Gamma_c$.
Here, we choose frequencies  $\omega_h$ and  $\omega_c$ as control parameters.
Note that it is not possible to optimize $E$ in Eq. (\ref{ef})
w.r.t both $\omega_c$ and $\omega_h$ simultaneously. Such two-parameter optimization   
yields the trivial solution, $\omega_c=\omega_h=0$.
Therefore, we will consider the optimization problem  w.r.t one parameter only, while keeping the other one 
fixed at some given value. First, keeping 
$\omega_h$ fixed, we optimize Eq. (\ref{ef}) w.r.t.
$\omega_c$, i.e., by setting $\partial E/\partial\omega_c=0$, we evaluate EMEF as
\begin{equation}
\eta^E_{\omega_h} = 1 + \frac{\tau}{\gamma}-\frac{\sqrt{(1+\gamma)\tau[\gamma+(2+\gamma)\tau]}}
{\sqrt{2}\gamma},  \label{effecofix1}
\end{equation}
Now,  $\eta^E_{\omega_h}$ is a monotonically increasing function of
$\gamma$. Therefore, we can obtain the lower and upper bounds of EMEF by letting 
$\gamma\rightarrow 0$ and $\gamma\rightarrow\infty$, respectively. Further, writing in terms of $\eta_{\rm C}^{}=1-\tau$, we have
\begin{equation}
\frac{3}{4}\eta_{\rm C}^{} \leq \eta^E_{\omega_h} \leq 1 - \sqrt{\frac{(1-\eta_{\rm C}^{})(2-\eta_{\rm C}^{})}{2}}. \label{B1}
\end{equation}
The lower bound, $3\eta_{\rm C}^{}/4$, obtained here is also derived as the lower bound for low-dissipation heat 
engines \cite{deTomas2013} and
minimally nonlinear irreversible heat engines \cite{LongLiu2014}. The upper bound,
$\sqrt{(1-\eta_{\rm C}^{})(2-\eta_{\rm C}^{})/2}$,  
was first derived by Angulo-Brown for a classical endoreversible heat engine \citep{ABrown1991}. 
Henceforth, we denote it as 
$\eta_{\rm AB}$. Under the conditions of tight-coupling and symmetric dissipation, 
$\eta_{\rm AB}$ can also be obtained 
for low-dissipation heat engines \cite{deTomas2013} and minimally nonlinear irreversible heat engines \cite{LongLiu2014}. 

Alternately, we may fix the value of $\omega_c$ and optimize 
$E$ w.r.t $\omega_h$, thus obtaining EMEF in the following form 
\begin{widetext}
\begin{equation}
\eta^E_{\omega_c} = \frac{\gamma(1-\tau^2) + 2(1+\gamma)\tau-\sqrt{(1+\gamma)\tau^2(1+\tau)[\gamma+(2+\gamma)\tau]}}
{\gamma + 2\tau +3\gamma \tau}.   \label{effecofix2}
\end{equation}
\end{widetext}
Again $\eta^E_{\omega_c}$  is monotonic increasing function of
$\gamma$. So we obtain lower and upper bounds on EMEF in the limiting cases 
$\gamma\rightarrow 0$ and $\gamma\rightarrow\infty$, respectively. In terms of
$\eta_{\rm C}^{}$, we have
\begin{figure} [ht]
 \begin{center} 
\includegraphics[width=8.6cm]{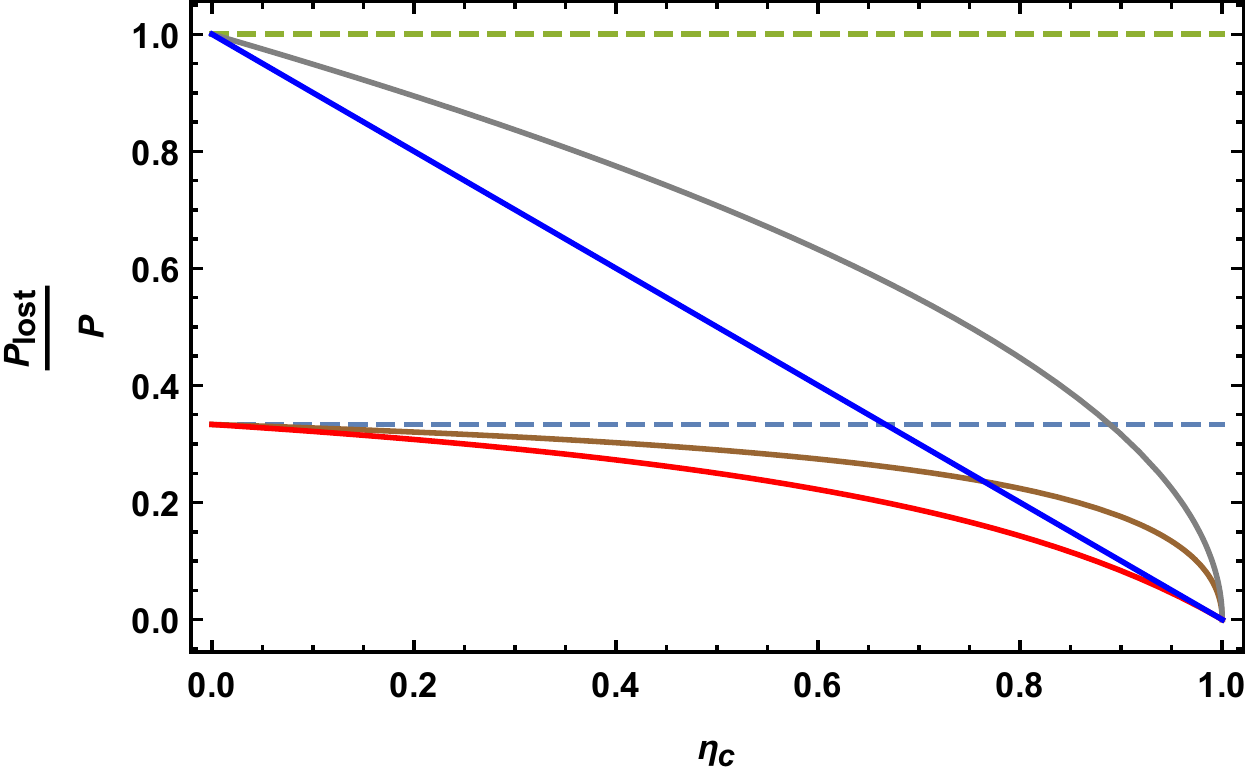}
 \end{center}
\caption{(Color online) Comparison of the ratios of power lost to power output, versus $\eta_{\rm C}^{} = 1-\tau$,  for the optimization 
of two different target functions - Ecological function and power output. The low lying set of curves 
[Eqs. (21) and (22)] represent the case of optimal EF, whereas the upper 
 curves [Eqs. (23) and (24)] represent the case of optimal power output.}
\end{figure}
\begin{equation}
\eta_{\rm AB} \leq   \eta^E_{\omega_c} \leq \frac{3-2\eta_{\rm C}^{}}{4-3\eta_{\rm C}^{}}\eta_{\rm C}^{}.\label{B}
\end{equation}
Under the conditions of extreme dissipation, upper bound $(3-2\eta_{\rm C}^{})\eta_{\rm C}^{}/(4-3\eta_{\rm C}^{})$ reported here, also serves as 
the  upper bound for the low-dissipation \cite{Esposito2010} and minimally non-linear 
irreversible \cite{LongLiu2014} heat engines.
\section{Fractional loss of power at maximum ecological function and maximum power output}
In this section, we compare performance of the three-level heat engine operating at maximum EF to
the engine operating at maximum power. 
Now, in the definition of EF,   
$T_c\dot{S}_{\rm tot}\equiv P_{\rm lost}$ represents 
the loss of power. So, we rewrite EF as 
$E=P-P_{\rm lost}$ and after rearranging terms, we obtain
\begin{equation}
R' \equiv \frac{P_{\rm lost}}{P} = 1 - \frac{E}{P} \equiv 1 - R. \label{Rdash}
\end{equation} 
We calculate $R$, and hence $R'$, 
in four different cases, as discussed in Appendix C. 
For optimization of EF w.r.t
$\omega_c$, at a fixed $\omega_h$, the ratio of optimal EF, 
$E^{*(\omega_h)}_{\rm eco}$, 
to the power at maximum EF, $P^{*(\omega_h)}_{\rm eco}$, 
is given by Eq. (\ref{rwheco}). We mention here
only the limiting cases $\gamma\rightarrow 0$ and  $\gamma\rightarrow \infty$, for which the respective 
equations for  $R'$ can be derived using Eqs. (\ref{Rdash}) and (\ref{rwheco2}): 
\begin{equation}
R'^{ \omega_h}_{{\rm eco}(0)} = \frac{1}{3},  \quad   R'^{\omega_h}_{{\rm eco}(\infty)} =
 \frac{\sqrt{\tau}}{\sqrt{\tau}+\sqrt{2(1+\tau)}}.
\end{equation}
Similar equations for the optimization of $E$ w.r.t $\omega_h$, while keeping $\omega_c$ fixed,
are given by
\begin{equation}
R'^{ \omega_c}_{{\rm eco}(0)} = \frac{\sqrt{\tau}}{\sqrt{\tau}+\sqrt{2(1+\tau)}}, \quad
R'^{\omega_c}_{{\rm eco}(\infty)} = \frac{\tau}{1+2\tau}.
\end{equation}
All the above expressions approach the value 1/3 near equilibrium, i.e. for small temperature differences. The fractional loss of power is, in general, higher
for the case with fixed $\omega_h$ than with a fixed $\omega_c$. 
As $\gamma$ increases, the fractional loss of power decreases. Also
note that $ R'^{\omega_h}_{{\rm eco}(\infty)}=R'^{ \omega_c}_{{\rm eco}(0)}$, as expected, since efficiencies are
also equal for the corresponding cases, 
$\eta^E_{\omega_h(\infty)}=\eta^E_{\omega_c(0)}=\eta_{\rm AB}$, as can be seen from Eqs. (\ref{B1}) and (\ref{B}).

Next we calculate the ratio of power loss to power output for the cases when we optimize power output. First, we 
discuss the case when the optimization is performed over $\omega_c$. As
seen from Eq. (\ref{zeropower}), $R^{ \omega_h}_{{\rm pow}(0)}=0$, which indicates that corresponding EF
is zero in this case, which in turn implies that the loss of power is equal to the power output. The ratios
$R'$ for the extreme cases $\gamma\rightarrow 0$ and $\gamma\rightarrow \infty$ are given by
\begin{equation}
R'^{ \omega_h}_{{\rm pow}(0)} = 1, \quad R'^{ \omega_h}_{{\rm pow}(\infty)} = \sqrt{\tau}
\end{equation}
The corresponding expressions, at optimal power output w.r.t $\omega_h$, are given by
\begin{equation}
R'^{ \omega_c}_{{\rm pow}(0)} =  R'^{ \omega_h}_{{\rm pow}(\infty)}, \quad R'^{ \omega_c}_{{\rm pow}(\infty)} =  \tau.
\end{equation}
Similar trend is observed for the fractional loss of power in this case also, 
as noted for the optimal EF above. More importantly, for near equilibrium conditions (small values of $\eta_C$), optimal EF yields  
lower values of fractional loss of power as compared to optimal power 
output. 

\section{Ratio of power at maximum ecological function to maximum power}
Fig. 3 indicates that the fractional loss of power is larger when the three-level laser heat engine 
operates at maximum power as compared to the engine operating at maximum EF. So, it is useful to evaluate
the  ratio of power output at maximum EF to the maximum power.  Defining
$  \bar{R}^{\omega_h}_{\gamma} =   P^{*(\omega_h)}_{\rm eco}/P^{*(\omega_h)}_{\rm pow}$ 
(see Eqs. (\ref{peco1}) and (\ref{power7})), 
and by taking limits $\gamma\rightarrow 0$ and $\gamma\rightarrow \infty$, we have following two equations, 
respectively
\begin{figure}   
 \begin{center}
\includegraphics[width=8.6cm]{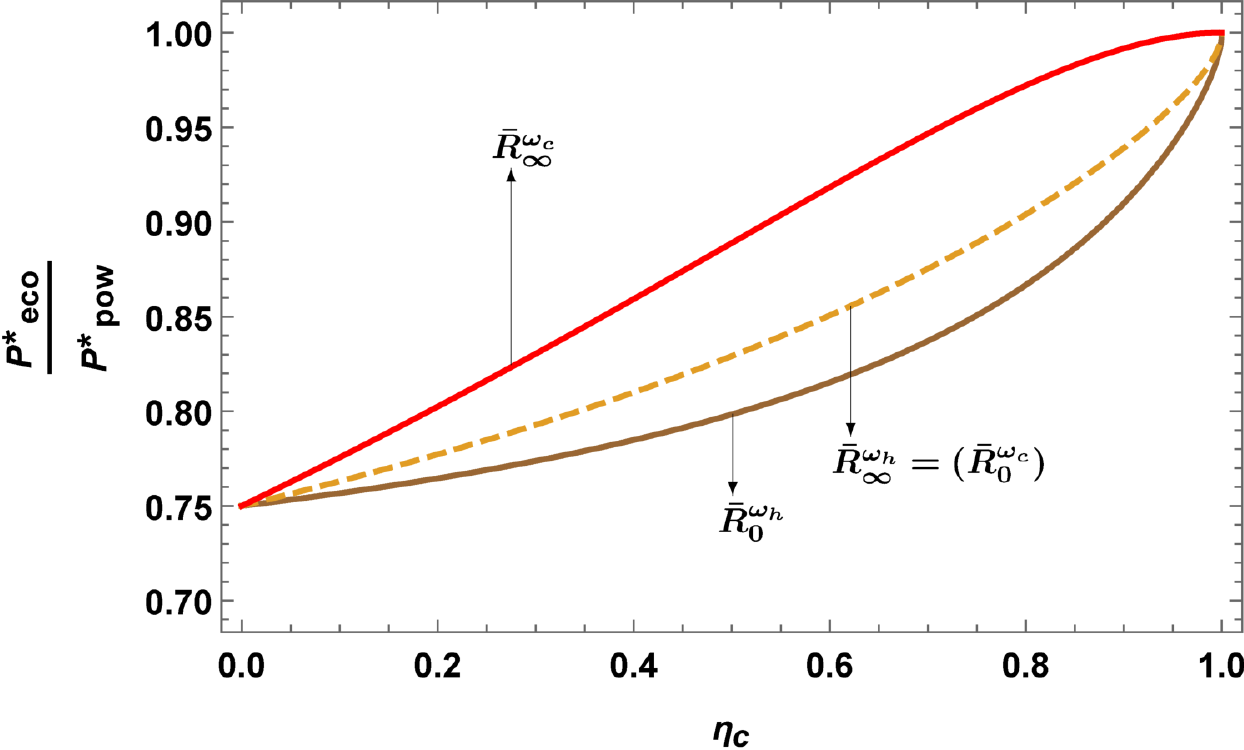}
 \end{center}
\caption{(Color online) Comparison of the ratio $\bar{R}$ of the power output at maximum EF 
to the maximum power [Eqs. (\ref{ratiop1}) and (\ref{ratiop2})].}
\end{figure}
\begin{equation}
\bar{R}^{\omega_h}_{0} = \frac{    1+3\tau-\frac{\tau(3+5\tau)}{\sqrt{1+3\tau}}   } {1+3\tau-2\sqrt{2\tau(1+\tau)}}
,
\bar{R}^{\omega_h}_{\infty} = \frac{  1+\tau -\frac{\tau(3+\tau)}{\sqrt{2\tau(1+\tau)}}   } {(1-\sqrt{\tau})^2}
\label{ratiop1}
\end{equation}
\begin{figure}   
 \begin{center}
\includegraphics[width=8.6cm]{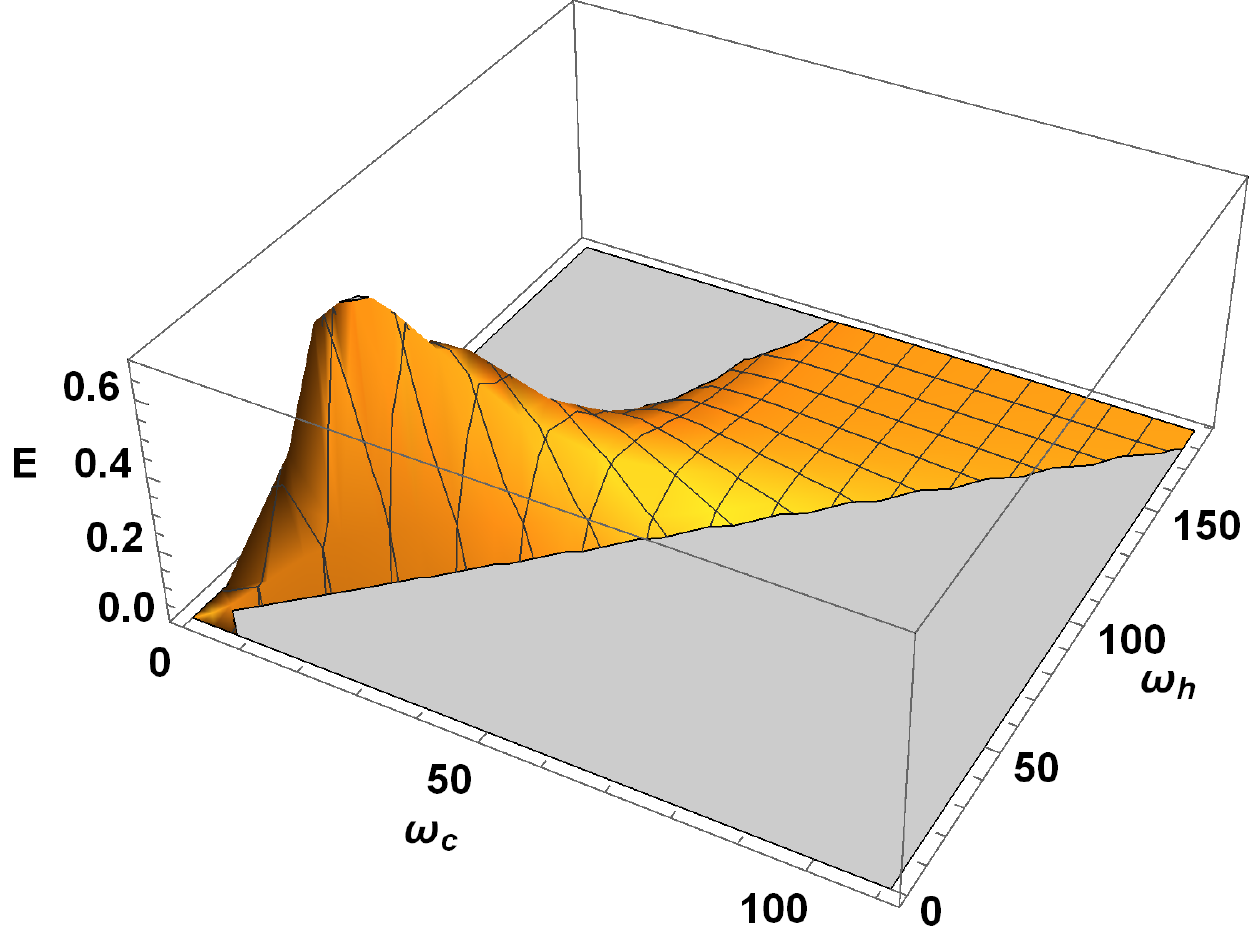}
 \end{center}
\caption{(Color online) 3D-plot of EF [Eq. (\ref{eco5}]) in terms of control frequencies $\omega_c$
and $\omega_h$ for $\hbar=1, k_{\rm B}=1, \Gamma_h=\Gamma_c=1, \lambda=1, T_h=20, T_c=5$.}
\end{figure}
Similar equations can be obtained for fixed $\omega_c$ by dividing 
$P^{*(\omega_c)}_{\rm eco}$ with $P^{*(\omega_c)}_{\rm pow}$ (see Eqs. (\ref{peco2}) and (\ref{power88})) 
and repeating the above mentioned step. Thus we have
\begin{equation}
\bar{R}^{\omega_c}_{0} = \bar{R}^{\omega_h}_{\infty},
\quad
\bar{R}^{\omega_c}_{\infty} = \frac{1+2\tau}{(1+\tau)^2}. \label{ratiop2}
\end{equation}
Again $\bar{R}^{\omega_h}_{\infty}$ is equal to $\bar{R}^{\omega_c}_{0}$ as expected. Plotting Eqs. (\ref{ratiop1})
and (\ref{ratiop2}) in Fig. 4, we observe that at least 75\% of the maximum power is produced 
in the maximum EF regime. The ratio $P^*_{\rm eco}/P^*_{\rm pow}$  increases with increasing 
$\eta_{\rm C}^{}$, which is expected since the efficiency of the engine also 
increases, while the dissipation decreases.
\section{Concluding Remarks}
We have analyzed and optimized thermodynamic performance of SSD heat engine with the ecological function (EF). Here we 
performed one parameter optimization of EF alternatively w.r.t $\omega_c$ ($\omega_h$ fixed) and 
$\omega_h$ ($\omega_c$ fixed)  and 
obtained the general expressions for the EMEF in high temperatures regime. In the limit of extremely asymmetric dissipation, 
lower and upper bounds on the efficiency are obtained.  $\eta_{\rm AB}$ serves as the upper bound in the former case and lower 
bound in the later case, thus separating the entire parameter regime of $\eta^E$ into two parts. To this end, we want to 
remark that although the two-parameter global maximum of EF does not exist under strong matter-field coupling ($\lambda \gg  \Gamma_{h,c}$) and in high temperatures limit, for the general case---where $\lambda$ may be comparable to $\Gamma_{h,c}$, numerical results indicate that the global maximum of EF  may exist (See Fig. 5). But, it is difficult to obtain analytic expressions for EMEF in this case. 

Finally, we have compared the 
performance of a quantum heat engine operating at maximum EF to the engine operating at maximum power. It is inferred that the fractional loss of power is appreciably low in case of engine operating at maximum EF while it 
produces at least 75\% of the power output by the engine working
in maximum power regime.  These conclusions concur with the optimal performance of a classical endoreversible heat engine operating in the ecological regime \cite{ABrown1991}. 
Hence, we conclude that classical as well as quantum heat engines  operating at maximum ecological function are much
more efficient and environment friendly than the engines operating at maximum power. Therefore, it is  reasonable as well as sensible to design real heat engines
along the lines of maximizing the ecological function, both for economical and ecological purposes.
Similarly, the analogue of ecological function for refrigerators 
\cite{Yan_1996, VarinderJohal} may be employed for quantum models where usually 
the cooling power is optimized \cite{Alonso2014B,Erdman_fridge}.  

\section*{Acknowledgements}
V. S. gratefully acknowledges insightful discussions with Professor Robert Alicki. The preliminary results of
the above paper were presented at the International conference "Workshop on Quantum and Nano Thermodynamics (WQNT 2018)" held at Alvkarleby, Sweden, for which 
financial support provided by Indian Institute of Science Education and Research, Mohali, India, is gratefully acknowledged.   
\appendix\section{Steady state solution  of density matrix equations}
Here, we solve the equations for density matrix in the steady state. Substituting the expressions for $H_0$, $\bar{H}$,
$V_0$, and using Eqs. (\ref{dissipator1}) and (\ref{dissipator2}) in Eq. (\ref{dm1}), the time evolution
of the elements of the density matrix are governed by following equations:
\begin{eqnarray}
\dot{\rho}_{11} &=& i\lambda (\rho_{10}-\rho_{01}) - 2\Gamma_h[(n_h+1)\rho_{11}-n_h\rho_{gg}],\label{A1} \\
\dot{\rho}_{00} &=& -i\lambda (\rho_{10}-\rho_{01}) - 2\Gamma_c[(n_c+1)\rho_{00}-n_c\rho_{gg}], \\
\dot{\rho}_{10} &=& -[\Gamma_h(n_h+1)+\Gamma_c(n_c+1)]\rho_{10} + i\lambda(\rho_{11}-\rho_{00}), \nonumber
\\
\\
\rho_{11} &= & 1-\rho_{00} - \rho_{gg}, \\
\dot{\rho}_{01} &=& \dot{\rho}_{10}^*. \label{A5}
\end{eqnarray}
Solving Eqs. (\ref{A1}) - (\ref{A5}) in the steady state by setting $\dot{\rho}_{mn}=0$ ($m,n=0,1$), we obtain
\begin{widetext}
\begin{equation}
\rho_{10} = \frac{i\lambda(n_h-n_c)\Gamma_c\Gamma_h}
{
\lambda^2[(1+3n_h)\Gamma_h + (1+3n_c)\Gamma_c] + \Gamma_c\Gamma_h[1+2n_h+n_c(2+3n_h)][(1+n_c)\Gamma_c + (1+n_h)\Gamma_h ] 
},\label{rho10}
\end{equation}
\end{widetext}
and 
\begin{equation}
\rho_{01} = \rho_{10}^*. \label{rho01}
\end{equation}
Calculating the trace in Eq. (\ref{power1}), the output power is given by
\begin{equation}
P = i\hbar\lambda(\omega_h-\omega_c)(\rho_{10}-\rho_{01}). \label{power3}
\end{equation}
Similarly evaluating the trace in Eq. (\ref{heat1}), heat flux $\dot{Q}_h$ can be written as
\begin{equation}
\dot{Q}_h = -\hbar \omega_h (2\Gamma_h[(n_h+1)\rho_{11}-n_h\rho_{gg}]). \label{heat3}
\end{equation}
Using the steady state condition $\dot{\rho}_{11}=0$ (see Eq. (\ref{A1})), Eq. (\ref{heat3}) becomes
\begin{equation}
\dot{Q}_h = -i\hbar \lambda\omega_h(\rho_{10}-\rho_{01}).
\end{equation}
Now EF is given by
\begin{equation}
E =  2P - (1-\tau) \dot{Q}_h. \label{eco3} 
\end{equation}
Using Eqs. (\ref{power3}) and (\ref{heat3}), we recast Eq. (\ref{eco3}) as follows
\begin{equation}
E =  i\hbar\lambda(\rho_{01}-\rho_{10})[2(\omega_h-\omega_c) - (1-\tau)\omega_h]. \label{eco4b}
\end{equation}
Substituting Eqs. (\ref{rho10}) and (\ref{rho01}) in Eqs. (\ref{power3}) and (\ref{eco4b}), we have
\begin{widetext}
\begin{equation}
P =    \frac{2\hbar\lambda^2 \Gamma_c\Gamma_h(n_h-n_c) (\omega_h-\omega_c) }
{
\lambda^2[(1+3n_h)\Gamma_h + (1+3n_c)\Gamma_c] + \Gamma_c\Gamma_h[1+2n_h+n_c(2+3n_h)][(1+n_c)\Gamma_c + (1+n_h)\Gamma_h] 
},
\label{peco}
\end{equation}
\begin{equation}
E = \frac{2\hbar\lambda^2 \Gamma_c\Gamma_h(n_h-n_c)[2(\omega_h-\omega_c) - \eta_c\omega_h]}
{
\lambda^2[(1+3n_h)\Gamma_h + (1+3n_c)\Gamma_c] + \Gamma_c\Gamma_h[1+2n_h+n_c(2+3n_h)][(1+n_c)\Gamma_c + (1+n_h)\Gamma_h]}.
\label{eco5}
\end{equation}
\end{widetext}
\section{Optimization in the high temperatures limit}
In order to obtain analytic expressions of interest, we optimize power output and the EF given above, in the high temperatures limit, while assuming a  strong matter-field coupling $\lambda \gg  \Gamma_{h,c}$. 
In the said limit, $n_h$ and $n_c$ can be approximated as
\begin{eqnarray}
n_h &=& \frac{1}{e^{\hbar\omega_h/k_B T_h}-1}\simeq \frac{k_B T_h}{\hbar\omega_h}, \label{nh}
\\
n_c &=& \frac{1}{e^{\hbar\omega_c/k_B T_c}-1}\simeq \frac{k_B T_c}{\hbar\omega_c}. \label{nc}
\end{eqnarray}
Using Eqs. (\ref{nh}) and (\ref{nc}) in Eq. (\ref{eco5}),  and ignoring the terms containing $\Gamma_{h,c}$ in 
comparison to $\lambda$, we can write $P$ and $E$ in terms of $\tau=T_c/T_h$ and $\gamma=\Gamma_h/\Gamma_c$ in the
following form
\begin{eqnarray}
P &\simeq& \frac{ 2\hbar\Gamma_h(\omega_h-\omega_c)(\omega_c - \tau\omega_h )}
 {3(\omega_c \gamma + \tau\omega_h )},\label{power6} \\
E &\simeq& \frac{2\hbar \Gamma_h(\omega_c - \tau\omega_h)[2(\omega_h-\omega_c) - (1-\tau)\omega_h]}
 {3(\omega_c \gamma + \tau\omega_h )}.\label{eco6}
\end{eqnarray}
\subsection*{One parameter optimization of ecological function}
We optimize $E$ w.r.t to either $\omega_c$ or $\omega_h$, while keeping the other fixed. 
Optimizing Eq. (\ref{eco5}) w.r.t $\omega_c$, for a fixed $\omega_h$, and
solving for $\omega_c$, we obtain
\begin{equation}
\omega_c^*= \frac{\omega_h}{2\gamma} \Big(\sqrt{2\tau(1+\gamma)[\gamma+(2+\gamma)\tau]}-2\tau\Big). \label{omegac}
\end{equation}
Using Eq. (\ref{omegac}) in Eq. (\ref{efficiency}), EMEF is given by Eq. (\ref{effecofix1}).
Optimizing Eq. (\ref{eco6}) w.r.t $\omega_h$, we have
\begin{equation}
\omega_h^* = \omega_c \frac{ \sqrt{(1+\gamma)(1+\tau)[\gamma+2(1+\gamma)\tau]} -\gamma(1+\tau)}
{\tau(1+\tau)}.\label{omegah}
\end{equation}
In this case, EMEF is given by Eq. (\ref{effecofix2}).
\section{Ratio $E/P$ for two different target functions}
Here, we derive the expressions for the ratio $E/P$ 
for the following four cases.
\subsection*{Optimal $E$ for a fixed $\omega_h$}
The optimal value of the EF, $E^{*(\omega_h)}_{\rm eco}$,
can be evaluated by substituting Eq. (\ref{omegac}) into
Eq. (\ref{eco6}). Similarly, substituting Eq. (\ref{omegac}) into Eq. (\ref{power6}), we obtain the expression for power at 
maximum EF, $P^{*(\omega_h)}_{\rm eco}$. Therefore, we have
 \begin{eqnarray}
 E^{*(\omega_h)}_{\rm eco} &=& 
 \frac{2\hbar\omega_h\Gamma_h}{3\gamma^2}    \big(\gamma+4\tau+3\gamma\tau- 2 A\big), \label{ecoE1}
 \\
 P^{*(\omega_h)}_{\rm eco} &=& \frac{\hbar\omega_h\Gamma_h}{3\gamma^2 A} (A-2(\gamma+\tau))
 (A -2(1+\gamma)\tau), \label{peco1}
 \end{eqnarray}
where $A = \sqrt{2(1+\gamma)\tau[\gamma+(2+\tau)\gamma]}$. The ratio of 
$E^{*(\omega_h)}_{\rm eco}$ and $P^{*(\omega_h)}_{\rm eco}$ is evaluated to be
\begin{equation}
R^{\omega_h}_{\rm eco (\gamma)} = \frac{A}
{(1+\gamma)\tau + A}. \label{rwheco}
\end{equation}
Now, consider $\gamma\rightarrow 0$ and $\gamma\rightarrow \infty$. 
For these limiting cases, the above equation reduces to:
\begin{equation}
R^{\omega_h}_{{\rm eco}(0)} = \frac{2}{3},\quad R^{\omega_h}_{{\rm eco}(\infty)}
 =
 \frac{\sqrt{2(1+\tau)}}{\sqrt{\tau}+\sqrt{2(1+\tau)}}. \label{rwheco2}
\end{equation}
\subsection*{Optimal $E$ for a fixed $\omega_c$}
In this case, the expression for optimal EF and the corresponding 
power output can be obtained by using Eqs. (\ref{omegah}) and (\ref{eco6}), and Eqs. (\ref{omegah}) 
and (\ref{power6}), respectively.
 \begin{equation}
 E^{*(\omega_c)}_{\rm eco} = \frac{2\hbar\omega_c\Gamma_h \big( 1+3\tau+2\gamma(1+\tau)
 -2B \big)}{3\tau} ,\label{ecoE2}
 \end{equation}
\begin{widetext}
 \begin{equation}
 P^{*(\omega_c)}_{\rm eco} = \frac{2\hbar\omega_c\Gamma_h \big( (1+\gamma)(1+\tau) + B \big)
   \big( (\tau+\gamma)(1+\tau)+B\big)}
 {3\tau(1+\tau)B} \label{peco2},
 \end{equation}
 \end{widetext}
 where $B=\sqrt{(1+\gamma)(1+\tau)[\gamma+(2+\tau)\gamma]}$.
We evaluate the ratio of $E^{*(\omega_c)}_{\rm eco}$ and $P^{*(\omega_c)}_{\rm eco}$ as follows
\begin{equation}
R^{\omega_c}_{\rm eco (\gamma)} = \frac{B}
{(1+\gamma)\tau + B }.
\end{equation}
Again the limiting cases $\gamma\rightarrow 0$ and $\gamma\rightarrow \infty$ yield the following two 
equations
\begin{equation}
R^{\omega_c}_{{\rm eco}(0)} = \frac{\sqrt{2(1+\tau)}}{\sqrt{\tau}+\sqrt{2(1+\tau)}},         
\quad 
		R^{\omega_c}_{{\rm eco}(\infty)} = \frac{1+\tau}{1+2\tau}. 
\end{equation}
\subsection*{Optimal power with a fixed $\omega_h$}
The optimization of power w.r.t $\omega_c$ ($\omega_h$ fixed) or $\omega_h$ ($\omega_c$ fixed) is 
perfomed in the Ref. \cite{Dorfman2018}. 
For the former case, the expression for $\omega_c^{P*}$ is given by
\begin{equation}
\omega_c^{P*} = \gamma^{-1} [\tau + \sqrt{\tau(1+\gamma)(\tau+\gamma)}]\omega_h. \label{omegacP}
\end{equation}
Again, the expressions for optimal power and for the EF at optimal power 
are evaluated to be 
 \begin{eqnarray}
 P^{*(\omega_h)}_{\rm pow} &=& \frac{2\hbar\omega_h\Gamma_h  (  \gamma +2\tau + \gamma\tau 
 - 
 2 C  )}{3\gamma^2} \label{power7},
\\
 E^{*(\omega_h)}_{\rm pow} &=& \frac{2\hbar\omega_h\Gamma_h (C -(1+\gamma)\tau ) 
 (C-2\tau -(1+\tau)\gamma )  } 
 {\gamma^2 C} \label{eco7} \nonumber ,
 \\
 \end{eqnarray}
% %
where $C=\sqrt{\tau(1+\gamma)(\tau+\gamma)}$.
The required ratio is calculated to be 
\begin{equation}
R^{\omega_h}_{\rm pow (\gamma)} = 1 - \frac{C}{\tau+\gamma}, \label{ratio7}
\end{equation}
from which we can write 
\begin{equation}
R^{\omega_h}_{{\rm pow}(0)} = 0, \quad R^{\omega_h}_{{\rm pow}(\infty)} = 1-\sqrt{\tau}.\label{zeropower}
\end{equation}
\subsection*{Optimal power with a fixed $\omega_c$}
For this case, the optimal value of  $\omega_h^{P*}$ is given by \cite{Dorfman2018}
\begin{equation}
\omega_h^{P^*} = \tau^{-1} [-\gamma + \sqrt{(1+\gamma)(\tau+\gamma)}]\omega_c. \label{omegahP}
\end{equation}
Then, we can obtain
\begin{eqnarray}
 P^{*(\omega_c)}_{\rm pow} &=& \frac{2\hbar\omega_c\Gamma_h \big(1+\tau+2\gamma-2 D \big)} {3\tau},  \label{power88}
\\
 E^{*(\omega_c)}_{\rm pow} &=& \frac{ \big(D -(1+\gamma)  \big)
 \big( (1+\tau)D-2\tau -\gamma(1+\tau) \big) }
 {(2\hbar\omega_c\Gamma_h)^{-1} 3\tau \sqrt{(1+\gamma)(\tau+\gamma)}},\nonumber
 \\
\end{eqnarray}
where $D=\sqrt{(1+\tau)(\tau+\gamma)}$. The ratio of 
$E^{*(\omega_c)}_{\rm pow}$ to $P^{*(\omega_c)}_{\rm pow}$ is evalauted to be 
\begin{equation}
R^{\omega_c}_{\rm pow (\gamma)} = 1-  \frac{(1+\gamma)\tau}    {D}, 
\end{equation}
whose limiting values 
for $\gamma\rightarrow 0$ and $\gamma\rightarrow \infty$, are
\begin{equation}
R^{\omega_c}_{{\rm pow}(0)} = 1-\sqrt{\tau}, \quad   R^{\omega_c}_{{\rm pow}(\infty)} = 1-\tau.
\end{equation}
%

% \bibliography{biblo}

\bibliographystyle{apsrev4-1}

\end{document}